# Human and Machine as Seen at the Co-Creation Age: A Co-Word Analysis in Human Machine Co-creation (2014–2024)

**Mengyao Guo[1], Jinda Han[2], Ze Gao[3], Yuan Zhuang[4], Xingting WU[5*]**

[1]Harbin Institute of Technology, Shenzhen, China
[2]University of Illinois at Urbana Champaign, Champaign, USA
[3]Cyanpuppets, Guangzhou, China
[4]Shandong University, Jinan, China
[5]Swinburne University of Technology, VIC, Australia

**ABSTRACT**

This paper explores the evolving landscape of human-machine co-creation, focusing on its development in the context of the ACM Conference on Human Factors in Computing Systems (CHI) from 2014 to 2024. We employ co-word analysis to identify emerging trends, central themes, and the intellectual trajectory of this field. The study highlights the shift from viewing machines as mere tools to recognizing them as collaborative partners in creative processes. By understanding these dynamics, we aim to provide insights into the implications of this paradigm shift for creativity, innovation, and societal impact, ultimately fostering a more inclusive and effective approach to human-machine interaction in various domains.

**Keywords:** Human machine co-creation, Design, Co-design, HCI, Co-word analysis

## INTRODUCTION

In late 1968, the Museum of Modern Art showed a groundbreaking exhibition titled "The Machine as Seen at the End of the Mechanical Age." Running from November 1968 to February 1969, this pivotal showcase, curated by K.G. Pontus Hultén, marked the culmination of one machine era while unknowingly standing at the threshold of another (Pontus Hultén, 1968). The exhibition reflected on the mechanical age - a period characterized by gears, levers, and visible moving parts. It celebrated the artistic and cultural impact of machines that had shaped the Industrial Revolution and modern society. However, even as visitors contemplated these mechanical marvels, a new machine age was quietly dawning. As Hegel says, "The owl of Minerva takes flight only at dusk (Houlgate, 2005)." The flight of this wise owl began in the twilight of the mechanical age and opened the dawn of a new machine era - the age of computers and artificial intelligence (AI).

Over half a century later, we find ourselves fully immersed in this new machine era - one defined by computers and artificial intelligence (Tegmark, 2018). They are fundamentally rooted in physical hardware with functional intangible or "digital" cause the computer, which is a physical machine of tremendous complexity, serves as the foundation for all of our modern computational and AI capabilities. This transition from the mechanical age to the computer-based age represents a profound shift in our relationship with machines (traditional and modern machines) (Brynjolfsson and McAfee, 2014). Where once we interact with machines through visible mechanisms, we now engage with computers through interfaces, data, and binary. Yet, behind every algorithm, every AI model, and every instance of human machine interaction lies the physical reality of computer hardware, including processors, memory, and storage devices - our new machine.

In this context, we turn our attention to a phenomenon that epitomizes this brand-new mechanical age: Human Machine Co-Creation. This concept represents a paradigm shift from viewing machines as mere tools to seeing them as collaborative entities in the creative process (Barile et al., 2024). It embodies the potential of combining human creativity with the computational power and capabilities of modern computers and AI systems. As we explore the field of human machine co-creation today, we carry forward the spirit of reflection embodied in that seminal 1968 MoMA exhibition. Just as Hultén and his contemporaries examined the impact of mechanical machines on art and society, we now seek to understand the implications of computers and AI on our creative endeavors.

To shed light on this evolution, we employ co-word analysis (Cambrosio et al., 1993), a method that investigates the simultaneous appearance of keywords across various publications. This method enables us to delineate the intellectual landscape of human machine co-creation, revealing emerging trends, core themes, and shifting paradigms. Prior studies have highlighted the significance of keyword analysis in assessing the contributions across different fields, such as children's computer interaction (Giannakos et al., 2020), accessibility research in HCI (Sarsenbayeva et al., 2023), intellectual progress in the ACM Conference on Human Factors in Computing Systems (CHI) (Liu et al., 2014), identifying both limitations and potential avenues for growth. Such investigations illustrate that analyzing the terminology within a research area can uncover its conceptual framework and developmental path.

This comprehensive survey examines the intellectual landscape of human machine co-creation from 2014 to 2024, a period marked by significant technological advancement and paradigm shifts. We anchor our analysis in the proceedings of the ACM CHI, acknowledging its pivotal role in the HCI community for three key reasons: its rigorous peer-review process ensures research quality; its position as a bellwether for emerging trends in human-computer interaction; its consistent documentation of technological innovations. This methodological decision enables us to construct a representative overview of the field's development.

## RELATED WORK

Polese et al. (2022), in their research on human machine interactions for value co-creation, investigate the role of human machine interactions in enhancing decision-making processes. They contribute to the field by proposing a shift from AI to intelligence augmentation (IA), emphasizing the collaborative integration of human and machine intellectual processes to generate a positive differential in decision-making outcomes. The study highlights the importance of understanding value co-creation processes within service ecosystems, suggesting that humans and machines should be considered actors in these ecosystems to improve AI design and human training for effective and trustworthy decision-making. However, the research is still in its infancy, with weak connections established between human machine interaction, decision-making, and service ecosystem viability. Their research inspires us to explore how human machine co-creation can go further and in which areas to derive different insights for users and developers.

While in Zhuo's work (2021), he explores the innovative concept of human machine co-creation in the realm of artistic paintings, proposing a three-stage model that categorizes machine involvement as expression tools and mediums of experience. The primary contribution of this work is the development of a collaborative framework where machines assist artists in exploring new painting strategies, enhancing creativity, and improving efficiency through computational experiments and robotic assistance. A notable limitation is the nascent stage of machine involvement, suggesting that the full potential of such collaborations is yet to be realized. This research can inspire further studies by demonstrating how machines can transform traditional artistic processes, offering new avenues for creativity and interaction and encouraging the exploration of deeper human machine partnerships in creative fields.

Yang et al. (2020) developed the complexities of designing effective human-AI interactions. They identify two primary sources of human-AI interaction design challenges: the uncertainty surrounding AI's capabilities and the complexity of AI outputs, which range from simple to highly adaptive systems. The authors synthesize existing literature and their own research experiences to map these challenges onto established design processes, such as the double diamond model, highlighting the difficulties faced by both user-centered and technology-driven design approaches. They propose a framework categorizing AI systems into four levels, each presenting distinct design challenges, and demonstrate its utility for designers, researchers, and tool makers in addressing these issues. The paper also emphasizes the evolving nature of AI as a design material, which complicates traditional HCI and UX design methods, and suggests that understanding these complexities can lead to more effective human-AI interaction designs. This work is supported by insights from workshops and critiques from HCI and machine learning researchers, underscoring the collaborative effort in refining the proposed framework.

## METHOD

To explore the changing landscape of research in human machine co-creation, we adopted a two-pronged methodological approach. Initially, we undertook a thorough process of collecting pertinent research articles from the CHI conference, which provided a solid and representative dataset for our analysis. Following this, we utilized co-word analysis to examine the connections between key themes and concepts within the gathered literature. This approach allowed us to uncover insights into the relationships that define this evolving field.

## Co-Word Analysis

To expand our understanding, we employ co-word analysis, a bibliometric method used to analyze textual content, focusing on the relationships between terms in context (Sedighi, 2016). This approach is particularly useful for mapping patterns and trends in scientific disciplines using publication data. The analysis is based on the premise that keywords appearing together in papers are likely related, with higher co-occurrence frequency suggesting stronger relationships.

In our analysis, we use two standard measures of density and centrality from graph theory (Cobo et al., 2011). Density measures the internal cohesion of a theme, with high density indicating closely related keywords within a cluster (Callon et al., 1991). Centrality measures a theme's importance in the overall research network, with high centrality suggesting the theme is crucial to the field's development (Callon et al., 1991).

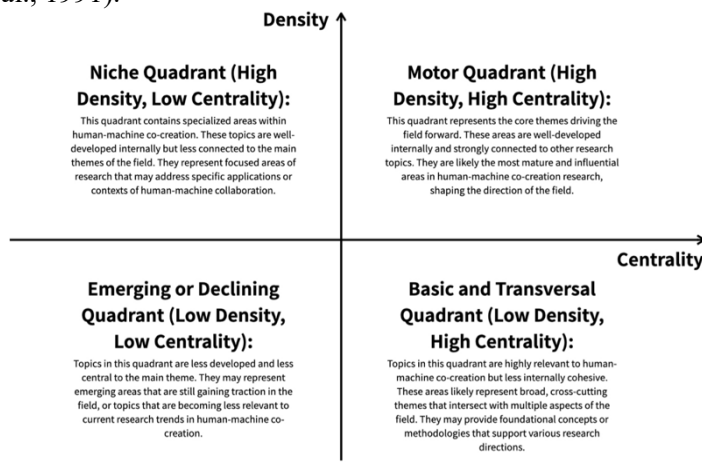

**Figure 1:** Strategic diagram of density and centrality based on Human Machine Co-Creation.

Our strategic diagrams are divided into four quadrants (see Figure 1): the Motor Quadrant (*high density, high centrality*) represents core themes driving the field forward; the Niche Quadrant (*high density, low centrality*) includes well-developed but less connected themes; the Emerging or Declining Quadrant (*low density, low centrality*) represents areas still gaining traction or losing relevance; and the Basic and Transversal Quadrant (*low density, high centrality*) contains themes highly relevant to human machine co-creation but less internally cohesive (Liu et al., 2014).

## Data Collection

To compile a relevant corpus for our analysis, we focused our search on CHI, a leading venue in the Human-Computer Interaction domain. We queried the ACM Digital Library for papers published between January 2014 and August 2024, which enabled us to identify significant trends and developments within the field. Utilizing a systematic filtering approach, we included only those papers that contained specific keywords in their titles, abstracts, or author-provided keywords: *human\** or *use\** (*e.g.*, human, humans, user-centered), *machin\** or *comp\** or *arti\** (*e.g.*, machine, computer, artificial intelligence), *co-creat\** (*e.g.*, co-creation, co-creative) and *creat\** (*e.g.*, creation, creativity, creative). These keywords were chosen because they cover a sizable enough spectrum of papers on human-machine co-creation and prevent us from choosing terms like 'collaborative robots,' or 'immersive design,' which could lead to bias. To match the previously mentioned search parameters, we relied on the advanced search feature of ACM Digital Library. For additional analysis, the returned results were exported into a different spreadsheet file (.BibTeX).

Our initial query returned 3,014 results related to Human Machine Co-Creation from the ACM Full-Text Collection across all available years in CHI Proceedings. To focus our analysis on recent trends, we narrowed our scope to papers published in the last eleven years; from January 2014 to August 2024, the ACM Digital Library recorded eleven CHI Conferences, and we got 1,784 research articles. We refined this dataset, which includes research articles rather than other types of submissions, such as posters, and workshop papers. This decision ensured our analysis was based on fully developed and peer-reviewed research. After applying these criteria, we conducted a manual review to ensure the relevance and appropriateness of the papers in our dataset. During this process, we identified and excluded several articles that did not align with our focus on CHI proceedings. We exclude nine articles from the Proceedings of the Fifth/Sixth Workshop on Beyond Time and Errors: Novel Evaluation Methods for Visualization, two articles from the Proceedings of HCI Korea, and one article from the Proceedings of the 1st ACM SIGCHI International Workshop on Investigating Social Interactions with Artificial Agents. These exclusions were necessary to maintain the integrity of our dataset and ensure that our analysis accurately reflected trends within the CHI conference proceedings. After these refinements, our final dataset comprised 1,772 research articles.

From the 1,772 articles, we extracted all author-assigned keywords (N = 5,302) from the papers and manually revised and grouped them under a unified overarching common keyword, e.g., keywords 'gpt', 'chatgpt', 'gpt-2', 'gpt-3' and 'gpt4' were grouped into 'chatgpt.' Keywords appearing in singular and plural form, spelling, abbreviations, and acronyms were also merged. We further established a frequency threshold ($f \geq 8$) and removed fewer assigned keywords when analyzing our dataset, e.g., 'animal computer interaction,' 'Amazon echo,' and 'brainwriting.' Three of the paper's authors manually and collaboratively scanned through the keywords and grouped the keywords, resulting in a total of 139 unique keywords used in our in-depth analysis.

## RESULTS

The CHI proceedings data (1,772 research articles) over the past eleven years show distinct phases of growth, a dramatic spike, and sustained expansion in publications, which can be attributed to the growing popularity of human machine co-creation (see Figure 2). Publications slightly increased from 53 to 71 between 2014 and 2016, indicating early research interest. With 223 publications in 2017, there was a notable outlier that may indicate momentary increased interest. Following a brief dip, publications grew year over year from 2018 to 2024, surpassing the 2017 peak by 2023 (227 publications) and reaching 274 in 2024. This trajectory shows that within the HCI community, research on human machine co-creation is growing rapidly and becoming more and more relevant.

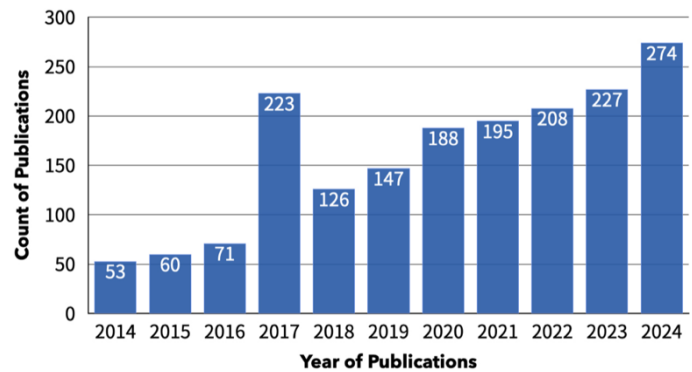

**Figure 2:** Number of publications per year.

## Thematic Cluster

Our thematic clustering analysis of 139 high-frequency keywords in human machine co-creation research employed a novel two-stage approach: Large Language Model (LLM)-based clustering followed by human expert refinement.

In the first stage, we utilized an LLM (Claude-3.5-Sonnet) to perform semantic clustering. The model analyzed each keyword within the context of human machine co-creation, computing semantic similarities and applying clustering algorithms to group-related terms. This approach provided an efficient initial structure for our analysis, identifying potential semantic relationships among the keywords. The second stage involved human expert intervention to refine and optimize the results. Experts validated LLM-generated clusters in three steps. They first checked each keyword's contextual significance within its cluster to find thematically discordant words. Second, they resolved semantic ambiguity by examining each keyword's many interpretations and categorizing it according to the study setting. Finally, they manually reassigned misclassified keywords using the thematic framework and concept links. This iterative review approach modified the thematic framework to better reflect data patterns while keeping classification system logic. Our method resulted in the identification of ten distinct clusters (see Table 1) and their keyword numbers / total frequency in four quadrants (see Figure 3): Artificial Intelligence and Machine Learning (A), Data and Privacy (B), Health and Assistive Technologies (C), Human-Computer Interaction Design (D), Innovation and Creativity (E), Internet of Things and Smart Devices (F), Social and Collaborative Technologies (G), Special Topics and Emerging Fields (H), User Research and Methodologies (I), and Virtual and Augmented Reality Technologies (J).

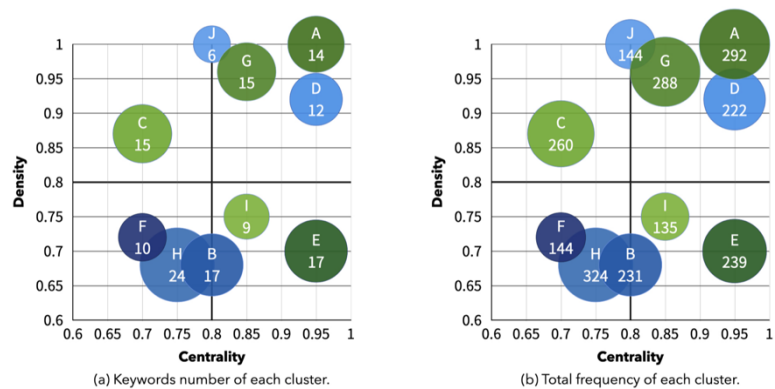

**Figure 3:** The keyword numbers and total frequency of each cluster.

**Table 1:** Research Themes of 2014-2024: size, total frequency (TF), centrality (Centr.), density (Dens.).

| ID | Keywords in each Cluster (2014–2024) | Size | TF | Centr. | Dens. |
|---|---|---|---|---|---|
| A | algorithms, artificial intelligence, chatgpt, computer vision, explainable ai, generative ai, human-ai collaboration, human-ai interaction, human-centered ai, large language models (llms), machine learning, natural language processing, learning, automation | 14 | 292 | 0.95 | 1.00 |
| B | data, data visualization, personal data, personal informatics, privacy, security, sensemaking, surveillance, transparency, visual analytics, personalization, trust, visualization, ethics, fairness, infrastructure, policy | 17 | 231 | 0.80 | 0.68 |
| C | accessibility, affective computing, assistive technology, autism, blindness, dementia, disability, health, healthcare, mental health, visual impairment, wellbeing, blind, covid-19, emotion | 15 | 260 | 0.70 | 0.87 |
| D | critical design, design, design methods, design research, human-computer interaction (hci), interaction, interaction design, soma design, speculative design, user experience, user-centered design, human-robot interaction | 12 | 222 | 0.95 | 0.92 |
| E | 3d printing, creativity, creativity support, creativity support tool, digital fabrication, diy, making, personal fabrication, prototyping, rapid prototyping, design fiction, fabrication, craft, diversity, participation, performance, decision making | 17 | 239 | 0.95 | 0.70 |
| F | internet of things (iot), mobile, mobile applications, smart home, smartphone, smartwatch, ubiquitous computing, wearables, smart textiles, home | 10 | 144 | 0.70 | 0.72 |
| G | co-design, collaboration, community, computer supported cooperative work (cscw), computer-mediated communication (cmc), crowdsourcing, online communities, participatory design, remote collaboration, social interaction, social media, social support, twitter, misinformation, open source | 15 | 288 | 0.85 | 0.96 |
| H | behavior change, children, conversational agent, crisis informatics, education, embodied interaction, embodiment, feminist hci, game design, games, gender, gestures, haptics, older adults, robot, storytelling, sustainability, sustainable hci, voice assistants, chatbot, communication, social robots, youth, india | 24 | 324 | 0.75 | 0.68 |
| I | ethnography, field study, hci4d, interviews, literature review, qualitative study, research through design, survey, user study | 9 | 135 | 0.85 | 0.75 |
| J | augmented reality, avatar, mixed reality, presence, virtual reality, wearable computing | 6 | 144 | 0.8 | 1.00 |

This approach, combining computational semantics with domain expertise, yielded a robust and meaningful categorization of keywords in human machine co-creation research, as evidenced by the coherent groupings in the final dataset.

## Keyword Clusters Analysis

This analysis explores the core developments, health technologies, privacy concerns, and smart device integration in human machine co-creation, revealing key trends and potential future directions in the field.

## Core Development of Human Machine Co-Creation

### Cluster A: Artificial Intelligence and Machine Learning

This cluster, positioned in the Motor Quadrant with high centrality (0.95) and density (1.00), including 14 keywords, is a primary driver of innovation in human machine co-creation. The high frequency of keywords such as "artificial intelligence" (58 occurrences) and machine learning (49 occurrences) underscores their fundamental importance in the field. There are some other notable trends within this cluster. The rise of "large language models (llms)" (37 occurrences) and "generative ai" (26 occurrences) indicates a shift towards more sophisticated, content-generating AI systems. Increasing focus on "human-ai collaboration" (22 occurrences) and "human-AI interaction" (21 occurrences), suggesting a move towards more symbiotic relationships between humans and AI. Emerging interest in "explainable ai" (9 occurrences) and "human-centered ai" (10 occurrences), reflecting growing concerns about AI transparency (Felzmann et al., 2020) and usability (Chen et al., 2021). This cluster reveals a field that is not only advancing technically but also becoming more attuned to human factors and ethical considerations (Eigenstetter, 2020). The presence of "natural language processing" (12 occurrences) and "computer vision" (11 occurrences) indicates that AI is being applied to a wide range of human-like cognitive tasks (Gonzalez, 2024).

### Cluster D: Human-Computer Interaction Design

The Human-Computer Interaction Design cluster has 15 keywords, it is occupied in the Motor Quadrant with high centrality (0.95) and density (0.92). This cluster emphasizes the crucial role of design in human machine co-creation (Barile et al., 2024; Fu and Zhou, 2020; Lataifeh et al., 2024; Zhu et al., 2018). The prominence of "design" (55 occurrences) as a keyword underscores its central importance. In this cluster, we find a strong focus on a user-centered approach to technology development from the "user experience" (28 occurrences) and "interaction design" (21 occurrences) (Battistoni et al., 2023; Bond et al., 2019; Margetis et al., 2021; Stige et al., 2023). The presence of "speculative design" (17 occurrences) and "critical design" (8 occurrences) suggests an interest in exploring future possibilities and challenging existing paradigms in HCI (Alfrink et al., 2023; Carvalho et al., 2022; Grba, 2022; Jang and Nam, 2022). The inclusion of "human-robot interaction" (13 occurrences) indicates the extension of HCI principles to physical robotic systems (Bartneck et al., 2024; Selvaggio et al., 2021).

This cluster reflects a field that is deeply committed to creating intuitive, effective, and meaningful interactions between humans and machines (Cross and Ramsey, 2021; Nardo et al., 2020; Schuetz and Venkatesh, 2020). The presence of "design research" (17 occurrences) and "design methods" (9 occurrences) suggests a rigorous, methodological approach to advancing the field (Dellermann et al., 2021; Seeber et al., 2020; Yuk et al., 2022).

### Cluster E: Innovation and Creativity

This cluster is located in the Basic and Transversal Quadrant with high centrality (0.95) but lower density (0.70) with 17 keywords, this cluster highlights the creative aspects of human machine co-creation (Fu and Zhou, 2020; Kantosalo et al., 2021; Wu et al., 2021). It shows a strong focus on physical creation, as evidenced by keywords such as "fabrication" (22 occurrences) (Bickel et al., 2023; Scott and Ali, 2021), "3D printing" (20 occurrences) (Chaudhuri et al., 2023), and "digital fabrication" (17 occurrences) (Song, 2020; Soomro et al., 2021). It emphasizes "creativity" (17 occurrences) and "creativity support" (11 occurrences), indicating interest in how technology can enhance human creative processes. While the presence of "design fiction" (21 occurrences) suggests an interest in speculative and future-oriented design approaches (Cox, 2021; Ghajargar et al., 2022; Ringfort-Felner et al., 2022).

This cluster's composition reflects a trend toward integrating creative and maker practices in human machine co-creation. The inclusion of "prototyping" (18 occurrences) and "rapid prototyping" (16 occurrences) indicates a focus on iterative, hands-on development processes (Chaudhuri et al., 2023; Votintseva, 2023; Xiong et al., 2023).

**Cluster G: Social and Collaborative Technologies**

This cluster, positioned in the Motor Quadrant with high centrality (0.85) and density (0.96), this cluster emphasizes the social aspects of technology in human machine co-creation (Sawaragi et al., 2020). In this cluster, the dominance of "social media" (51 occurrences) as a keyword reflects its significant concentration on media in human machine co-creation (Adikari et al., 2021). It also shows a strong focus on collaborative approaches, as evidenced by "co-design" (39 occurrences), "crowdsourcing" (37 occurrences) (Geyer et al., 2021; van Rijn et al., 2024), and "participatory design" (36 occurrences) (Frauenberger et al., 2016; Pearson et al., 2017). The inclusion of "computer supported cooperative work (cscw)" (11 occurrences) (Martinez-Maldonado et al., 2020; O'Toole, 2023) and "computer-mediated communication (cmc)" (11 occurrences) (Liu et al., 2022; Park et al., 2022) indicates ongoing interest in how technology facilitates human collaboration (Fui-Hoon Nah et al., 2023).

This cluster's composition reflects a trend towards more inclusive, participatory, and socially aware approaches in HCI. The presence of "misinformation" (11 occurrences) (Kabir et al., 2024) as a keyword suggests that researchers are also addressing the challenges and potential negative impacts of these social technologies (Younes and Al-Zoubi, 2015).

**Cluster J: Virtual and Augmented Reality Technologies**

It is in the Motor Quadrant with high centrality (0.80) and density (1.00), this cluster focuses on immersive technologies. Despite having fewer keywords (6 keywords), its position indicates the significant role of these technologies in shaping the future of human machine interaction. The notable trends include: the dominance of "virtual reality" (60 occurrences) (Dufresne et al., 2024; Shen et al., 2024) and "augmented reality" (38 occurrences) (Cheng et al., 2023; Liu et al., 2023), indicating strong research interest in immersive experiences; the inclusion of "mixed reality" (22 occurrences) (Johnson et al., 2023; Schlagowski et al., 2023) suggests a focus on blending virtual and physical environments (Chollet et al., 2009; Kantosalo et al., 2021); the presence of "presence" (8 occurrences) (Dufresne et al., 2024; Newhart and Olson, 2017; Venkatraj et al., 2024) as a keyword indicates interest in the psychological aspects of immersive experiences.

This cluster's composition reflects the growing importance of immersive technologies in human machine co-creation, suggesting a trend towards more engaging and immersive digital experiences.

**Health Technologies and Emerging Domains**

**Cluster C: Health and Assistive Technologies**

Cluster C represents a well-developed but specialized niche within human machine co-creation research with 15 keywords. Its low centrality (0.70) but high density (0.87) suggests a focused area of study that may be somewhat isolated from other research domains. The cluster's emphasis on "accessibility" (59 occurrences) (Cha et al., 2024; Li et al., 2023; Tang et al., 2023), "mental health" (27 occurrences) (Cuadra et al., 2024; Hwang et al., 2024; Yoo et al., 2024), and specific health conditions such as "visual impairment" (15 occurrences) (Herskovitz et al., 2023) and "dementia" (12 occurrences) (Marchetti et al., 2022) indicates a strong commitment to leveraging technology for improving health outcomes and quality of life. The inclusion of "affective computing" (13 occurrences) (Murali et al., 2021; Wan et al., 2024) and "emotion" (10 occurrences) (Kim et al., 2024) suggests a trend toward more empathetic and responsive health solutions. The presence of "COVID-19" (12 occurrences) as a keyword demonstrates the field's agility in addressing urgent global health challenges. This cluster reflects a human-centered approach to health technology, aiming to create inclusive and targeted solutions for diverse health needs.

**Cluster H: Special Topics and Emerging Fields**

Cluster H (Special Topics and Emerging Fields) has 24 keywords, and it embodies the dynamic and evolving nature of human machine co-creation research. Its position in the Emerging/Declining Quadrant with low centrality (0.75) and density (0.68) indicates that these topics are not yet fully integrated into the core of the field but represent potential areas for future growth. The cluster's diversity is striking, covering topics from age-specific technologies, "children" (35 occurrences) (Chen et al., 2024b; Dwivedi et al., 2024; Rocha et al., 2023) and "older adults" (18 occurrences) (Harrington and Egede, 2023) to "sustainability" (26 occurrences) (Crosby et al., 2023), "conversational agent" (24 occurrences) (Schmidt et al. 2022), and "embodied interaction" (13 occurrences) (Deppermann and Streeck, 2018; Montirosso and McGlone, 2020). This breadth suggests that researchers are exploring numerous avenues for expanding the scope of human machine co-creation, often intersecting with broader societal concerns. The presence of topics like "feminist hci" (12 occurrences) (Park et al., 2024; Søndergaard and Campo Woytuk, 2023) and "crisis informatics" (8 occurrences) (Soden et al., 2022) indicates a growing awareness of the social and ethical implications of technology development.

The interplay between Clusters C and H offers intriguing possibilities for the future of human machine cocreation research. The specialized health focus of Cluster C could benefit from the innovative approaches and broader perspectives represented in Cluster H. For instance, the emphasis on sustainability in Cluster H could inform the development of more environmentally conscious health technologies (Blandford, 2019). Similarly, the focus on embodied interaction and haptics in Cluster H could lead to more intuitive and engaging health interventions (Kelly et al., 2019). The attention to specific age groups in Cluster H complements the healthspecific focus in Cluster C, potentially leading to more targeted and effective health solutions across the lifespan. This intersection highlights the potential for cross-pollination between established health technology research and emerging HCI concepts (Holeman and Kane, 2020), potentially driving innovations that are technologically advanced and more inclusive, sustainable, and attuned to diverse human needs and societal challenges (Cutillo et al., 2020).

**Integrating Methods for Enhanced Privacy Protection**

**Cluster B: Data and Privacy**

Cluster B emphasizes the protection and ethical use of data in human machine co-creation. Key themes include "privacy" (42 occurrences) (H. Tan et al., 2022; Lee et al., 2024), "ethics" (20 occurrences) (Hanschke et al., 2024; Sparrow et al., 2024), "trust" (19 occurrences) (Harrington and Egede, 2023), and "data visualization" (23 occurrences) (Burns et al., 2023). This cluster reflects growing concerns about data privacy, personal data management, and the need for transparency in data-driven systems (Gorkovenko et al., 2020).

**Cluster I: User Research and Methodologies**

Cluster I focuses on approaches to understanding users and contexts. Prominent methods include "research through design" (34 occurrences) (Boucher, 2023; Gaver et al., 2022), "ethnography" (17 occurrences) (Chen et al., 2024a; Kang et al., 2022), and "literature review" (16 occurrences) (Liu et al., 2024). It also highlights hci4d (16 occurrences) (Kotut et al., 2020), indicating attention to diverse global contexts. This cluster represents a comprehensive approach to gathering user insights.

Integrating Clusters B and I significantly enhance data privacy research. User research methodologies from Cluster I could provide valuable insights into user perceptions and behaviors related to privacy (Chalhoub et al., 2020), informing more effective protection measures. Ethnographic approaches could reveal how users interact with data-driven systems in real contexts, while research through design could develop innovative approaches to data visualization and transparency. This integration would lead to more comprehensive, ethically sound, and user-friendly approaches to data privacy in human machine co-creation (Zheng et al., 2019).

**Smart Devices in Everyday Life**

**Cluster F: Internet of Things and Smart Devices**

Cluster F focuses on integrating connected technologies into daily life, emphasizing "internet of things (iot)" (38 occurrences) (Desjardins et al., 2020) and "wearables" (25 occurrences) (Olwal et al., 2020). It spans personal devices to "smart homes" (smart home, 13 occurrences; home, 11 occurrences) (Yao et al., 2019), highlighting "mobile applications" (8 occurrences) (Wardle et al., 2018) and emerging trends like "smart textiles" (8 occurrences) (Vogl et al., 2017). This user-centric approach aims to enhance experiences through seamless integration (Partarakis and Zabulis, 2024). The cluster reflects a dynamic field in human machine co-creation, balancing technological advancement with user needs and ethical considerations and underscoring the importance of interdisciplinary collaboration in IoT development.

## FINDING AND DISCUSSION

**Findings**

In corresponding to the keyword clusters analysis of core development of human machine co-creation, health technologies, and emerging domains, integrating methods for enhanced privacy protection and smart devices in everyday life, we divided our findings into redefining creativity, personalized health, privacy in the age of co-creation as well as everyday innovation to discuss, we aim to provide a comprehensive understanding of the dynamic interplay between human creativity and machine capabilities in contemporary contexts.

**Redefining Creativity:** The field is undergoing a profound transformation with the advancement of sophisticated AI systems, particularly in language models and generative AI. This shift is fundamentally altering the nature of human machine collaboration, pushing the boundaries of what's possible in co-creation (Woodruff et al., 2024). Large language models, for instance, are not just tools for natural language processing but are becoming active participants in creative processes, capable of generating complex narratives, code, and even conceptual ideas. This raises intriguing questions about the nature of creativity itself and the role of AI in augmenting human cognitive processes (Hassani et al., 2020; Markauskaite et al., 2022; Marrone et al., 2022).

The integration of advanced AI systems with user-centered design approaches is creating a new paradigm in human machine interaction (Margetis et al., 2021). We're moving beyond simple command-based interfaces to more intuitive, context-aware systems that can anticipate user needs and adapt in real-time. This symbiosis between human intuition and machine precision has the potential to unlock new forms of creativity and problem-solving that neither humans nor machines could achieve independently (Hancock, 2017; Jarrahi, 2018).

Moreover, the incorporation of fabrication technologies into the co-creation process is bridging the gap between digital conception and physical realization in unprecedented ways (Nechkoska et al., 2023). This convergence is not just about rapid prototyping; it's about creating a seamless continuum between thought, digital design, and physical manifestation. It challenges our traditional notions of the creative process and opens up possibilities for new forms of expression that blend the digital and physical realms in novel ways. The rise of immersive technologies like VR and AR in this context is particularly significant (Nussipova et al., 2020; Zhang et al., 2024). These technologies are not just new interfaces; they're entirely new environments for co-creation. They offer the potential to manipulate and interact with complex data and designs in three-dimensional space, potentially revolutionizing fields from architecture to scientific visualization. This immersive co-creation could lead to insights and innovations that are difficult to achieve in traditional 2D interfaces.

**Personalized Health:** In health technologies, human machine co-creation pushes the boundaries of personalized medicine and adaptive healthcare. The application of AI in this domain goes beyond mere data analysis; it's about creating intelligent systems that can learn from individual patient data, adapt to changing health conditions, and work in tandem with healthcare professionals to provide unprecedented levels of personalized care. For instance, developing AI-assisted diagnostic tools is about improving accuracy and creating systems that can explain their reasoning (Kumar et al., 2023), learn from human experts, and integrate complex, multifaceted health data in ways that humans alone might struggle to do (Chang, 2020). This could lead to a new paradigm in medical decision-making, where AI and human expertise work in synergy to provide more comprehensive and nuanced patient care.

**Privacy in the Age of Co-creation:** Integrating AI in human machine co-creation processes has brought privacy concerns to the forefront. As these systems become more sophisticated and pervasive, they collect and process vast amounts of personal data, raising critical questions about data ownership, consent, and the potential for misuse (Van den Hoven van Genderen, 2017). This challenge pushes researchers and designers to develop new approaches that embed privacy protection into the fabric of co-creation systems. Rather than treating privacy as an afterthought, there's a growing emphasis on "privacy by design" principles (Shehzadi, 2024; Yanisky-Ravid and Hallisey, 2019). This involves creating AI systems that inherently respect user privacy, minimizing data collection and processing while delivering powerful co-creation capabilities. In addition, there's an increasing focus on user empowerment and transparency. Co-creation systems are designed with intuitive interfaces that give users greater control over their data and more precise insights into its use. This shift towards user-centric privacy models is not just about compliance with regulations but about fostering a new ethos of responsible innovation in the AI era.

**Everyday Innovation:** The proliferation of the Internet of Things (IoT) devices and wearable technologies is extending the realm of human machine co-creation into everyday life. This trend transforms ordinary objects and environments into potential platforms for creative collaboration between humans and AI. Smart home systems, for instance, are evolving from simple automation to becoming active partners in home management and design (Aldrich, 2003; Garg and Cui, 2022). They're learning from user behaviors and preferences to co-create living spaces that adapt to inhabitants' needs and moods in real-time. In wearable technology, devices are moving beyond health tracking to become collaborative tools for personal expression and performance enhancement (Ferreira et al., 2021; Gandy et al., 2017). This ubiquity of co-creation opportunities democratizes access to advanced creative tools and resources. It's allowing individuals to engage in complex design and problem-solving tasks that were once the domain of specialists. However, it also raises important questions about the digital divide and technological literacy, as the benefits of these advancements may not be equally accessible to all.

As these technologies become more integrated into our daily routines, a growing need for thoughtful design that respects user autonomy promotes meaningful human control, and fosters genuine creativity rather than passive consumption of AI-generated content is needed; over-dependence on AI will erode personal agency.

**Division of Responsibilities, Redefinition of Human Roles**

The rapid advancement of AI is reshaping the distribution of responsibilities between humans and machines, leading to a dynamic redefinition of roles. This shift is not a simple transfer of tasks but a complex interplay of capabilities. Machines increasingly take on responsibilities involving rapid data processing, pattern recognition, and consistent execution. This allows humans to focus on areas that leverage uniquely human traits such as emotional intelligence, contextual understanding, ethical judgment, and creative ideation. However, these domains are not static; they evolve as AI capabilities expand.

Interestingly, this redistribution fosters a symbiotic relationship where humans and machines enhance each other's capabilities (Karnouskos, 2022; Noble et al., 2022). In creative fields, AI tools augment human creativity by generating initial ideas or variations, while humans retain the crucial role of curation and emotional resonance. In scientific research, AI's data processing complements human scientists' ability to form hypotheses and interpret results within broader contexts. This collaboration is leading to scenarios where decision-making responsibilities are shared. AI systems are often designed to leave critical decisions to humans, recognizing the importance of human judgment in complex situations (Korteling et al., 2021). Conversely, humans increasingly rely on AI for data-driven insights to inform their decisions (Akter et al., 2021; Ntoutsi et al., 2020). This mutual augmentation pushes the boundaries of what's possible in various fields, from healthcare to artistic expression.

Moreover, AI's development is compelling humans to reposition themselves in areas that remain uniquely human or require human oversight. This includes ethical decision-making, policy formation, interpersonal relationships, and creative direction. As AI takes over more routine tasks, there's a growing emphasis on developing human skills in creativity, critical thinking, and interdisciplinary synthesis – abilities currently beyond machine capabilities. This push for repositioning is not just about finding new niches for human work; it's about redefining the essence of human contribution in an AI-augmented world. It's prompting a reevaluation of education and skill development, focusing on cultivating adaptability and the uniquely human capacity for holistic understanding and ethical reasoning.

The evolving division of responsibilities creates a collaborative framework where humans and machines complement each other's strengths and compensate for weaknesses. This symbiosis is changing what we do and how we think about our roles and potential, pushing us to explore new frontiers of creativity and cognition in harmony with our artificial counterparts.

## LIMITATION

By analyzing papers from CHI, we describe the intellectual landscape of human machine co-creation within a specific context, which may limit the generalizability of our findings. While the CHI conference is a premier venue for human-computer interaction research, it primarily reflects trends and discussions within this community.

In future research, we plan to expand our scope to include related ACM conferences and journals, such as conferences include ACM Creativity & Cognition (C&C), ACM International Conference on Multimedia (MM), ACM Symposium on Virtual Reality Software and Technology (VRST), ACM SIGGRAPH and ACM SIGGRAPH Asia, ACM Conference on Designing

Interactive Systems (DIS), and journals include ACM Transactions on Computer-Human Interaction (TOCHI), ACM Computing Surveys (CSUR), among others. Additionally, our dataset is confined to articles published between 2014 and 2024, potentially overlooking earlier foundational work that could provide valuable context. Although our focus on keywords and co-word analysis is effective for mapping trends, it may oversimplify the complexities of the research landscape, leading to the omission of nuanced discussions and interdisciplinary connections. To address this, we intend to apply the power law principle to concentrate on the most significant areas of research by defining a threshold—such as capturing 80\% of the area corresponding to the top keywords—to identify the most meaningful terms. This approach aims to emphasize impactful, emerging, or niche topics while excluding less significant keywords in future analyses. Furthermore, the dynamic nature of the field means that human machine co-creation is continually evolving, and our analysis captures only a snapshot in time; thus, emerging trends may shift rapidly, necessitating ongoing research to keep pace with technological advancements and changes in user interactions.

## CONCLUSION

The evolution of human machine co-creation signifies a transformative shift in our relationship with technology, driven by advancements in artificial intelligence and interactive design. As we move deeper into this co-creation age, advanced technology integration enhances human creativity and redefines roles and responsibilities in collaborative processes. Our study highlights several key themes: the redefinition of creativity through sophisticated AI systems, the potential for personalized health solutions, the urgent need for privacy considerations, and the everyday innovation facilitated by IoT and wearable technologies. Each area underscores the importance of a human-centered approach, ensuring that technology empowers and augments human capabilities rather than replacing them. Moreover, the findings emphasize the necessity of ethical frameworks and design principles prioritizing user agency and privacy. As we navigate this rapidly changing landscape, fostering a collaborative spirit between humans and machines will be essential for unlocking new possibilities in creativity, health, and everyday life. The future of human machine co-creation lies in leveraging the strengths of both entities to create enriching, inclusive, and innovative experiences that reflect our diverse needs and aspirations.

*Example for Preparation of Your Own Paper* 19